\documentclass[journal=nalefd]{achemso} 
\usepackage[utf8]{inputenc}
\usepackage{amsmath} 
\usepackage{amssymb}
\usepackage{color}            
\usepackage{soul}           
\usepackage{graphicx}
\usepackage{subfig}
\usepackage[version=4]{mhchem}
\SectionNumbersOn

\sethlcolor{yellow}
\newcommand{\REVISION}{}

\setkeys{acs}{email = false}
 
\newcommand{\BESOsep}{$0.326_{0.312}^{0.341}$\,eV}
\newcommand{\SRV}{$1.17_{0.41}^{2.60}\times10^{6}$\,cm/s}
\newcommand{\pdoping}{$9.67_{5.08}^{17.76}\times10^{19}$\,cm$^{-3}$}
\newcommand{\diameter}{$76_{68}^{84}$\,nm}
\newcommand{\usedwires}{11,487}

\newcommand{\T}{$1984_{1741}^{2240}$\,K}
\newcommand{\ta}{$193_{112}^{347}$\,fs}
\newcommand{\pdeg}{$1.5\times10^{20}$\,cm$^{-3}$}
\newcommand{\slope}{$0.91_{0.90}^{0.92}$}

\newcommand{\tamx}{1\,ps}
\newcommand{\tauFit}{1.10$\pm$0.02\,ps}
\newcommand{\pturnp}{$2.3\times10^{20}$\,cm$^{-3}$}
\newcommand{\IQEmx}{$2\%$}

\title{Sub-Picosecond Carrier Dynamics Explored using Automated High-Throughput Studies of Doping Inhomogeneity within a Bayesian Framework}

\author{Ruqaiya Al-Abri}
\email{ruqaiya.al-abri@postgrad.manchester.ac.uk}
\author{Nawal Al Amairi}
\author{Stephen Church}
\affiliation{Department of Physics and Astronomy and the Photon Science Institute, University of Manchester, Oxford Road, Manchester, M13 9PL, United Kingdom}
\author{Conor Byrne}
\affiliation{Department of Chemistry, University of Manchester, Oxford Road, Manchester, M13 9PL, United Kingdom}
\author{Sudhakar Sivakumar}
\affiliation{Department of Physics and NanoLund, Lund University, Box 118, SE-221 00, Sweden}
\author{Alex Walton}
\affiliation{Department of Chemistry, University of Manchester, Oxford Road, Manchester, M13 9PL, United Kingdom}
\author{Martin H. Magnusson}
\affiliation{Department of Physics and NanoLund, Lund University, Box 118, SE-221 00, Sweden}
\author{Patrick Parkinson}
\affiliation{Department of Physics and Astronomy and the Photon Science Institute, University of Manchester, Oxford Road, Manchester, M13 9PL, United Kingdom}

\begin{document}

\section*{Abstract}
Bottom-up production of semiconductor nanomaterials is often accompanied by inhomogeneity resulting in a spread in electronic properties which may be influenced by the nanoparticle geometry, crystal quality, stoichiometry or doping. Using photoluminescence spectroscopy of a population of more than \REVISION{11,000} individual Zn-doped GaAs nanowires, we reveal inhomogeneity in, and correlation between doping and nanowire diameter by use of a Bayesian statistical approach. Recombination of hot-carriers is shown to be responsible for the photoluminescence lineshape; by exploiting lifetime variation across the population, we reveal hot-carrier dynamics at the sub-picosecond timescale showing interband electronic dynamics. High-throughput spectroscopy together with a Bayesian approach are shown to provide unique insight in an inhomogeneous nanomaterial population, and can reveal electronic dynamics otherwise requiring complex pump-probe experiments in highly non-equilibrium conditions.   

\section{Introduction}
Bottom-up growth of nanomaterials is widely established as a highly scaleable methodology, capable of producing electronic materials from nanometre to micrometre lengthscales\cite{Thiruvengadathan2013NanomaterialApproaches}. However, this capability is tempered by the sensitivity of growth to local conditions, and variation in precursor concentration, ratio, or temperature can lead to significant variation in yield or functional performance\cite{Al-Abri2021_ACS}. Crucially, both ensemble and single-element characterization are highly challenging for nanomaterials, creating a bottleneck for their exploitation; the former as it cannot measure inhomogeneity and the latter because measurement of a local region may not represent the whole sample size.\cite{Richman2009}.
GaAs nanowires (NWs) have been widely studied for optoelectronic applications;\cite{Dasgupta2014, parkinson2007transient, gallo2011, wang2018gaas, li2020optical} they can be produced with high crystal quality\cite{Joyce2008, dhaka2012high} and provide a facile route to heterostructure design based on decades of experience in planar material growth.\cite{hyun2013nanowire} However, a large surface-to-volume ratio and a relatively high surface recombination velocity of $5.4\times10^{5}$\,cm/s (for 50\,nm-thick GaAs NW\cite{joyce2013electronic}) give rise to low quantum efficiency of the emission\cite{chang2012electrical, jiang2012long}. Photoluminescence quantum efficiencies as low as $0.1\%$ have been reported for uncapped GaAs NWs \cite{parkinson2009carrier} compared to 50$\%$ for high-quality InP NWs.\cite{gao2014selective} 
For both emissive and photovoltaic applications,\cite{kim2021doping} it is crucial to maximize the radiative efficiency, therefore several approaches have been developed to improve radiative emission. These include heavy doping, to dramatically increase the radiative recombination rate,\cite{zhang2016recombination,alanis2018optical} and passivating the NW surface with a higher-bandgap capping layer to decrease the non-radiative emission originating from surface states.\cite{zhou2019epitaxial,couto2012effect} 

Measuring recombination rates provides a direct means to assess radiative and non-radiative processes\cite{Milot2015Temperature-DependentFilms}. At the shortest timescales relevant for high recombination rates, carrier cooling processes are often apparent\cite{Gierz2013SnapshotsGraphene,Bernardi2015AbGaAs,Bailey1990NumericalGaAs}. This is particularly crucial in photovoltaics as the efficiency of these devices depends on the electron-hole separation before recombination process.\cite{fast2020hot} To obtain a comprehensive understanding of the carrier pathways, electronic properties such as surface, radiative, and Auger recombination must be measured. However, inter-wire variation in geometry and doping results in an inherent spread and a potentially unknown distribution of the recombination values across a population which could mask systematic trends. 
Where surface processes dominate, geometric inhomogeneity will dominate; ensemble measurements will be biased towards larger nanowires or to higher quantum efficiency subsets, and are therefore unreliable to assess a given growth. A statistically rigorous analysis must be able to reflect the inhomogeneity of the material, informed by numerous measurements.
For this particular type of problem, the Bayesian methodology can be used to model a distribution in properties such as doping, diameter or surface recombination velocity, and is highly suited to determine unknown parameters and put limits on their spread.\cite{gabbard2022bayesian} The Bayesian approach is based on representing all model parameters by probability distributions, which is refined from a prior distribution -- representing knowledge of the system before the data is considered -- to a posterior distribution, using the fit of the model to the data.\cite{thrane2019introduction} Bayesian approaches have been demonstrated in a wide range of domains such as astrophysics,\cite{smith2020massively,dobigeon2007joint} sensor analysis,\cite{mirsian2019new} and modelling NW growth given experimental data.\cite{huang2010physics} The Bayesian approach is particularly fruitful for studying novel forms of existing materials, as it provides a framework to make use of prior measurements from earlier study while allowing new data to be used refine and update these values.

In this study, we demonstrate that automated high-throughput imaging and spectroscopy of a large population of single NWs with a range of diameters and doping can be used to investigate doping inhomogeneity, carrier cooling and recombination processes at the sub-picosecond timescale, without using pump-probe measurement which induces highly non-equilibrium populations, and crucially, with statistical confidence. 
Our high-throughput approach is less biased towards high-efficiency subsets of the NWs which might otherwise have led to ensemble measurements underestimating the true range of recombination velocities.  We show that for Zn-doped GaAs NWs with a median diameter of \diameter~, the surface recombination velocity has a median value of \SRV, (upper and lower limits represent the interquartile range), consistent with 5-13$\times10^{5}$\,cm/s as previously measured for Zn-doped GaAs NWs.\cite{darbandi2016measurement,joyce2013electronic,Joyce_2017_ACS} 
Hole densities are inhomogeneous across the population, with an asymmetric distribution of \pdoping~which is not significantly correlated with NW diameter; this demonstrates that other factors such as temperature or precursor availability during the growth process more strongly determine doping inhomogeneity. Significantly, by linking effective carrier temperature to effective recombination lifetime, we observe that high doping plays a critical role in hot band-edge emission, and therefore, emission temperature can be an accurate clock for understanding carrier dynamics in sub-picosecond timescale.

\section{Discussion and Results}
\subsection{Low Power Photoluminescence Measurements}
\begin{figure}
\includegraphics[width=0.5\textwidth]{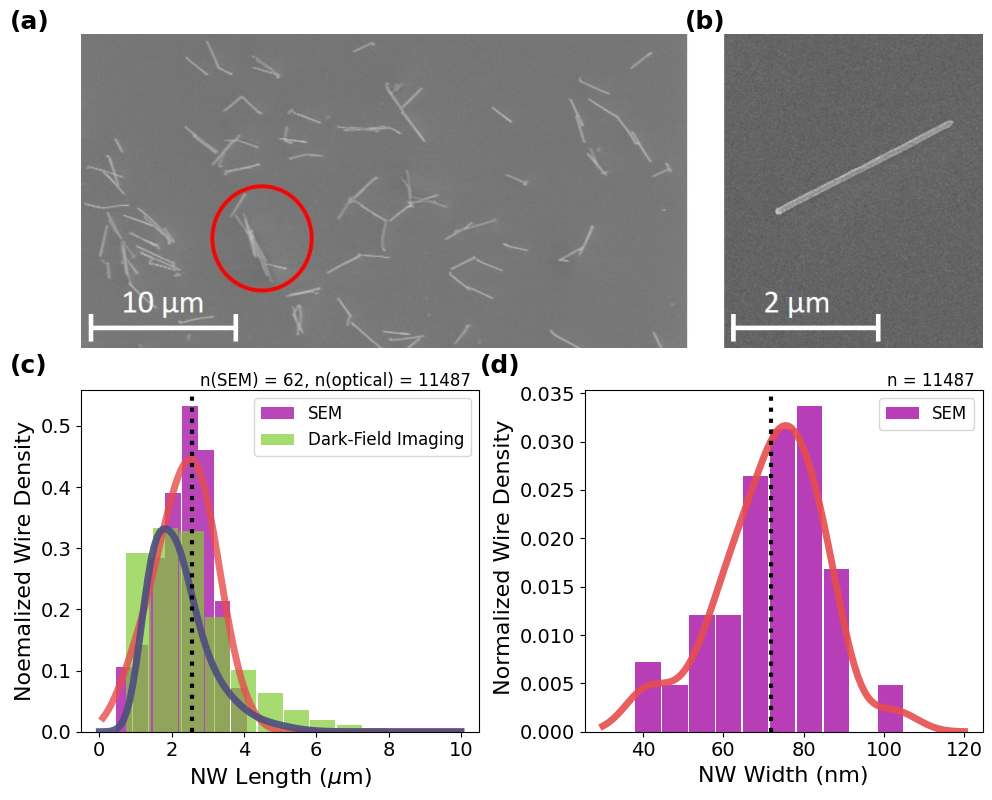}
\caption{(a) SEM image of an ensemble and (b) a single typical Zn-doped GaAs NW on silicon substrate. (c) Length distribution from SEM and filtered-NW length distribution from optical imaging. The vertical line indicates the median length from SEM (2.3\,$\mu$m). (d) Diameter distribution of the NWs obtained from the SEM, with the vertical line indicating the median diameter of the NWs from SEM (72\,nm). \REVISION{The numbers (n) at the top of (c) and (d) indicate the number of samples used to produce distributions.} Solid lines in c and d are kernel density estimates of the continuous probability distribution for \REVISION{the length and diameter data, respectively}.}
\label{fig:LengthWidth}
\end{figure}

Zn-doped GaAs NWs were grown using the Aerotaxy method\cite{heurlin2012continuous, Sivakumar2020Aerotaxy:Nanostructures} and were deposited on silicon with native oxide following a recipe described previously.\cite{yang2015zn} The NWs were heavily doped with atomic Zn with an ensemble average density of 2.32(14)$\times10^{21}$\,cm$^{-3}$ as measured using X-ray Photoelectron Spectroscopy (XPS) (details in the SI). A set of 20,000 NWs were initially located and investigated using automated micro-photoluminescence ($\mu$-PL) spectroscopy \cite{church2022holistic}. A continuous-wave HeNe laser of 632.8\,nm wavelength with circular polarization (to avoid polarization dependant absorption effects\cite{Titova2006}) and power density at the sample of 6.4\,kW cm$^{-2}$ was used (equivalent to around 30\,photons/picosecond/NW). Photoluminescence (PL) spectra and dark-field optical images were collected for each NW; the approximate length of each NW was extracted from the images.  
Scanning Electron Microscopy (SEM) was performed on the same sample to obtain a more accurate distribution of NW length and diameter from a subset of more than \REVISION{60}\,wires, with exemplary images shown in Figure \ref{fig:LengthWidth}. SEM images of a region with a high density of NWs and a typical single NW are illustrated respectively in Figure \ref{fig:LengthWidth}a and b. It is observed that a subset of wires studied may be clumps of multiple wires as shown circled in Figure \ref{fig:LengthWidth}a, due to high density of NW production by the Aerotaxy method.\cite{barrigon2017gaas} Analysis of the dark-field imagery and PL characteristics were used to exclude the majority of such objects, by filtering photoluminescence fit outside the expected range of energy, temperature or intensity as detailed in the SI. \REVISION{Initial threshold values were selected from a preliminary study of PL emission compared with dark-field imaging.} This approach eliminated clumps of wires (with high intensity) as well as surface contamination (with emission indistinguishable from background). Following this filtering process, around 46$\%$ of the initially identified objects were retained for study. 
This process may result in the removal of very weakly emitting single NWs; filtering produces a statistical bias which places a lower limit on the quantum efficiency of NWs that we can study. We address this through the incorporation of weighted evidence in our Bayesian model, discussed below. 
Figure \ref{fig:LengthWidth}c and d depict length and diameter distributions from SEM. For comparison, a distribution of length measured using optical microscopy obtained for the set of \usedwires~NWs used is shown in Figure \ref{fig:LengthWidth}c which is consistent with the SEM results. The NW diameter cannot be reliably determined using optical microscopy, as it is far below the optical resolution limit.

\subsection{Photoluminescence Modelling and Energy, Temperature, and Intensity Mapping}
\begin{figure}
\includegraphics[width=0.45\textwidth]{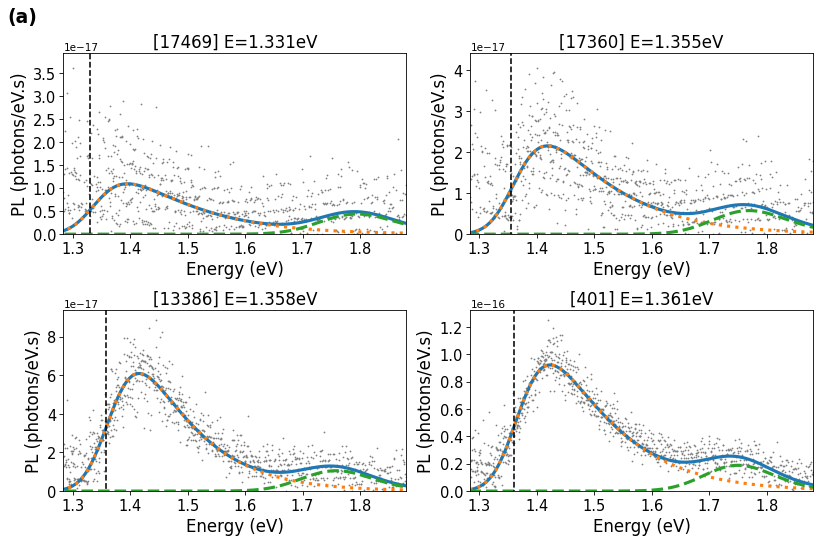}
\includegraphics[width=0.45\textwidth]{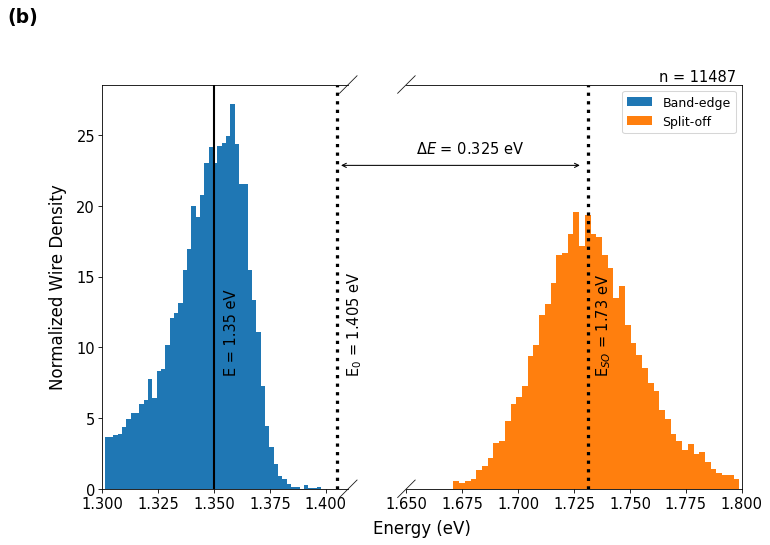}
\caption{(a) A series of PL spectra of Zn-doped GaAs NWs with fits as described in the text. The spectra are selected randomly and ordered by their redshift in band-edge emission. Best-fit parameter of the band-edge energy is given in the title, and the relative contribution of band-edge emission (orange dashed line) and split-off emission (green dashed lines) are shown. (b) Normalized distribution of the band-edge and split-off energies for \usedwires~NWs. \REVISION{The number (n) at the top of the figure indicates the number of samples used to generate the distribution.} The dotted vertical lines represent the updoped band-edge (1.405\,eV) and the median split-off band (1.73\,eV) energies of GaAs  at room temperature, \REVISION{the solid line represents the median doped band-edge}, and the horizontal line represents the separation between emission bands.}
\label{fig:SpectraFit}
\end{figure}

Previous photoluminescence studies with thicker NWs (d$>$200\,nm) showed a single emission peak at the band-edge,\cite{alanis2018optical} while studies of high-doped thin NWs (d$<$100\,nm) tend to show a higher energy peak\cite{yang2015zn} which has been associated with recombination from the conduction band to the split-off band with the transition energy around 0.33\,eV above that of the band-edge.\cite{benz1977auger,Olego1979,Ketterer2011} A series of randomly selected PL spectra for individual NWs are shown in Figure \ref{fig:SpectraFit}a. These spectra show an emission peak below 1.405\,eV attributed to red-shifted band-edge emission from GaAs, as well as a weaker second high-energy peak at around 1.73\,eV which we attribute to recombination from the conduction band to the split-off \REVISION{; which falls within the expected range for the split-off peak 1.65 to 1.9\,eV for Zn-doped GaAs}.\cite{Olego1980} \REVISION{Emission from conduction-band to split-off band recombination is often attributed to Auger population of the split-off band. In our case, we instead attribute this to direct absorption of light populating leading to creation of split-off band hole -- which has been noted for similar excitation energies at early times}\cite{Becker1988FemtosecondGaAs}\REVISION{ -- coupled with an  ultrashort carrier lifetime leading to all recombination processes completing within sub-picosecond timescales. As such, we neither expect not observe significant excitation density effects}.  

Each PL spectrum was fit using two models: one for the band-edge emission (BE) only, and one containing both band-edge and split-off peaks (BE+SO)\cite{alanis2017large}. For BE+SO model, the PL is fit with a linear sum of the two emissions,
\begin{align}
\label{eqn:model}
    I(E) &= \sum_{i=BE,SO}I_{i}(E) \\
     &= \sum_{i=BE,SO}\left[\beta_{i} (B(E,E_{g,i},T) \otimes G(E,\sigma_{i}))\right]
\end{align}
where each contribution is a convolution of $B$, the product of a three-dimensional density of states for a transition with energy $E_g$, and an occupation described by a Maxwell-Boltzmann distribution function with effective carrier temperature $T$
\begin{equation} \label{dos}
    B(E,E_g,T)  = \sqrt{(E-E_g)} e^{(-(E-E_g)/k_B T)},
\end{equation}
with a simple Gaussian distribution $G(E,\sigma)$
\begin{equation} \label{gaussian}
    G(E,\sigma) = \beta e^{(-E^2/2\sigma^2)},
\end{equation}
with width $\sigma$ representing all experimental sources of spectral broadening.  $\beta$ is a scaling factor corresponding to the intensity of each component in the spectrum. We justify using a Maxwellian temperature for emission, as we expect scattering and thermalization to occur on fast (sub-100\,fs) timescales in GaAs\cite{Taylor1985UltrafastCompounds, Bailey1990NumericalGaAs}. \REVISION{Figure} \ref{fig:SpectraFit}a \REVISION{shows a series of PL spectra with a best fit model. The full fitting parameters for these selected spectra are provided in the SI.} The quality of fitting was obtained for both models based on the reduced $\chi$-squared, and the most appropriate fit was selected for each spectrum for further analysis; around 1.5\% of spectra was better fit with band-edge emission only, and 98.5\% requiring two peak fits. The data and code used for this analysis are provided online.

Figure \ref{fig:SpectraFit}b shows the distribution in modelled emission energy ($E_g$) for the band-edge and the split-off emission. The band-edge energy shows a redshift with respect to intrinsic GaAs at room temperature (of 1.405\,eV indicated by a vertical line), with a median value of 1.34\,eV. This redshift is attributed to band-structure shift arising from heavy Zn doping.\cite{haggren2014effects,Alanis2019} The separation between the intrinsic band-edge energy and the median split-off band energy (\BESOsep) is consistent with reported split-off band separation for GaAs, within experimental uncertainty.\cite{benz1977auger,Zschauer1969AugerGaAs,olego1979luminescence}

The modelling above provides a number of parameters for each spectral fit, five of which provide insight into the recombination process; the band-edge emission energy $E_{g,BE}$, the effective carrier temperature for band-edge $T_{BE}$ and split-off emission $T_{SO}$, the band-edge emission amplitude $\beta_{BE}$ and split-off emission amplitude $\beta_{SO}$. The emission energy $E_{g,BE}$ has been shown to be strongly related to hole density for p-type material, and the doping level can be calculated based on the energy shift such that\cite{Borghs1989,alanis2018optical}

\begin{equation} \label{eqn:energy}
    E =  E_{0}-Kp^{1/3},
\end{equation}

where $E_{0}$ is the energy bandgap of intrinsic GaAs at room temperature, $p$ is the hole density, and $K$ is a constant determined for Zn-doped GaAs when using a 632.8\,nm excitation source that we have previously measured to be 1.158$\times10^{-8}$\,eV.cm using the same experimental system.\cite{Alanis2019} 

For very short emission lifetimes, we expect that carriers photogenerated with excess energy may not have cooled to the lattice temperature\cite{Aitchison1998EnhancedGaAs, Bernardi2015AbGaAs}. By approximating carrier cooling as a Newtonian process with a single effective cooling rate $\tau_0$, the carrier lifetime $\tau$ can be obtained from the carrier temperature $T$ for each NW
\begin{equation} \label{eqn:temperature}
    T = T_0 e^{(-\tau/\tau_0)} + T_L, 
\end{equation}
where $T_0$ is the initial temperature of electrons after photoexcitation with an upper limit given by the excess energy following the photon absorption process and $T_L$ is the lattice temperature. $\tau_0$ is the timescale of the dominant cooling process, and has been reported as approximately 0.2\,ps for \REVISION{undoped} GaAs \cite{yang2016observation} related to the longitudinal optical photon scattering rate.\cite{Scholz1998HolephononArsenide, Kash1989Carrier-carrierLuminescence} \REVISION{We expect early-time carrier cooling to be similar in our samples to that measured for undoped samples, where phonon-cooling is important process in both materials. While time-resolved techniques can -- in principle -- measure this cooling, our carrier lifetime of less than a picosecond and material volume of significantly less than 1\,$\mu$m$^3$ prohibits the straightforward application of such techniques.}
Using the emission temperature measured from PL as a proxy for carrier recombination lifetime is valid under certain conditions: that the recombination process takes place in the sub-picosecond timescale between thermalization and cooling, that low-density excitation is used to avoid carrier-carrier effects, and where a single effective cooling process dominates. In the modelling presented, we use the split-off band emission temperature to calculate the lifetime. In this case, the electron excess energy following photoexcitation is insufficient to populate the L or X valley, simplifying the interpretation as discussed later.

Finally, the total band-edge emission, given by the integral of the spectrum, can be used to understand the quantum efficiency of each NW. Internal quantum efficiency ($IQE$) is related to the $PL$ intensity by a constant experimental scaling factor $\alpha$ and a photon absorption rate $A(d)$, which varies with NW width $d$ to account for both absorption cross-section and absorption depth, 

\begin{equation} 
\label{eqn:pl}
    PL = \alpha A(d) {\rm{IQE}} = \alpha A(d) \left(\frac{Bp}{Bp+Cp^2+4S/d} \right), 
\end{equation}

where $Bp$, $Cp^2$, $4S/d$ are radiative, Auger, and surface recombination rates. The material parameters are: $B$ - the radiative recombination rate estimated between 10$^{-9}$ to 10$^{-10}$\,cm$^3$/s\cite{zhang2022all,strauss1993auger}, $C$ - the Auger rate estimated between 10$^{-26}$ to 10$^{-31}$ \,cm$^6$/s,\cite{capizzi1984electron,mclean1986picosecond} and $S$ - the surface recombination velocity estimated to be between 5-13$\times10^{5}$\,cm/s for diameter between 100-300\,nm.\cite{darbandi2016measurement, joyce2013electronic, Joyce_2017_ACS} The experimental scaling factor $\alpha$ is related to the conditions such as laser power and spot size, microscope collection efficiency, and spectrometer quantum efficiency. More details on the scaling $IQE$ to $PL$ is provided in the SI.

\subsection{Bayesian Modelling}
We propose a model for recombination that maps the doping $p$ and diameter $d$ for each NW to a unique triplet of observables, namely emission energy $E$ (through Equation~\ref{eqn:energy}), emission temperature $T$ (Equation~\ref{eqn:temperature}) and emission intensity PL (Equation~\ref{eqn:pl}). This mapping relies on a number of material, sample, and experimental-specific model parameters as listed in Table \ref{tab:priors}. These model parameters form a prior vector $\Psi$ which is used as an input within a Bayesian framework; they can be refined towards a posterior distribution by fitting the model to the data\cite{Trotta2008BayesCosmology, Bolstad2016}. \REVISION{These 10 parameters are linked to observables \textit{via} deterministic equations, however we emphasise that each equation is a functions of multiple Bayesian parameters. For example, $p$ is obtained from the emission peak $E$, however, the constants $E_0$ and $K$ are also Bayesian parameters which are allowed to vary within relatively tightly constrained distributions.} For every parameter, we define a prior -- a nominal probability distribution representing likely values -- using physical knowledge and literature values as summarized in Table \ref{tab:priors} (more details on prior notation is found in the SI). We sample from possible values of the parameters and make use of Eqns.~\ref{eqn:energy}, \ref{eqn:temperature} and \ref{eqn:pl} to produce a modelled three-dimensional distributions in $E'$, $T'$, and PL$'$. We performed Markov-Chain Monte-Carlo (MCMC) modelling with the Python \textit{emcee} package\cite{Foreman-Mackey2012Emcee:Hammer} to update the priors to maximize the probability of the model output given the experimental data - $P(\Psi | \rm{Data})$ - as expressed in Bayes formula
\begin{equation}
    P(\Psi | \rm{Data}) = \frac{P_{\rm{likelihood}}(\rm{Data} | \Psi) P_{\rm{prior}}(\Psi)}{P_{\rm{evidence}}(\rm{Data})}.
\end{equation}

\begin{table*}
\caption{Parameter description and prior distribution. Here $\mathcal{N}$, $G$, and $U$ donate normal, generalized normal, and uniform distributions (details in the SI). The mean is denoted by $\mu$, standard deviation $\sigma$, and shape parameter as $\beta$.\label{tab:priors}}
\resizebox{\textwidth}{!}{\begin{tabular}{|l|l|l|l|l|}
\hline
\textbf{Symbol}    & \textbf{Description}          & \textbf{Origin}          & \textbf{Prior}         & \textbf{Source}         \\ 
\hline \hline

E$_{0}$   & Energy band-edge of intrinsic GaAs [eV]             & Material        
& $\mathcal{N}$($\mu=1.405, \sigma=5e-3$)                                              & Ref[\citenum{kusch2014type}]\\ \hline
K         & Constant linking doping and redshift [eV.cm]            & Material        
& $\mathcal{N}$($\mu=1.158\times10^{-8}, \sigma=0.1\times10^{-8}$) and positive  & Ref [\citenum{alanis2018optical}]\\ \hline
log(B)    & Bimolecular radiative constant [cm$^3$/s]      & Material              
& $G$($\mu=-10, \sigma=1, \beta=8$) and $U$(-12,-8)                            & Ref [\citenum{nelson1978minority}]\\ \hline
log(C)    & Auger constant [cm$^6$/s]  & Material                                  
& $G$($\mu=-29, \sigma=1.5, \beta=8$) and $U$(-32,-26)                           & Ref [\citenum{ahrenkiel2001auger}]\\ \hline
log($\tau_{0}$)   & Initial cooling time [s]  & Material              
& $G$($\mu=-12.7, \sigma=0.5, \beta=8$) and $U$(-13.3,-11.5)    & Ref[\citenum{yang2016observation}]\\ \hline
log(p)    & Hole density in log$_{10}$-scale [cm$^{-3}$]          & Sample            
& $G$($\mu=20.4, \sigma=3.5, \beta=8$)                              & Ref[\citenum{Johansson2020CalculationNanowires}] \\ \hline
d         & NW diameter [nm]                                             & Sample                
& $\mathcal{N}$($\mu=77, \sigma=12.5$) and $U$(35,100)                                       &  SEM\\ \hline
log(S)   & Surface recombination velocity in log$_{10}$ [cm/s]          & Sample                
& $G$($\mu=6, \sigma=1.5, \beta=8$) and positive                            & Ref [\citenum{joyce2013electronic}]\\ \hline
T$_{0}$   & Initial temperature after excitation [K]            & Experimental          
& $G$($\mu=2000, \sigma=500, \beta=8$)                                                                     & Derived\\ \hline
log($\alpha$)  & Scaling factor related to experimental conditions on log$_{10}$ scale   & Experimental     
& $G$($\mu=-15.8, \sigma=2, \beta=8$)                                                                & Calculated\\ \hline
\end{tabular}}
\end{table*}

\begin{figure}
\includegraphics[width=0.38\textwidth]{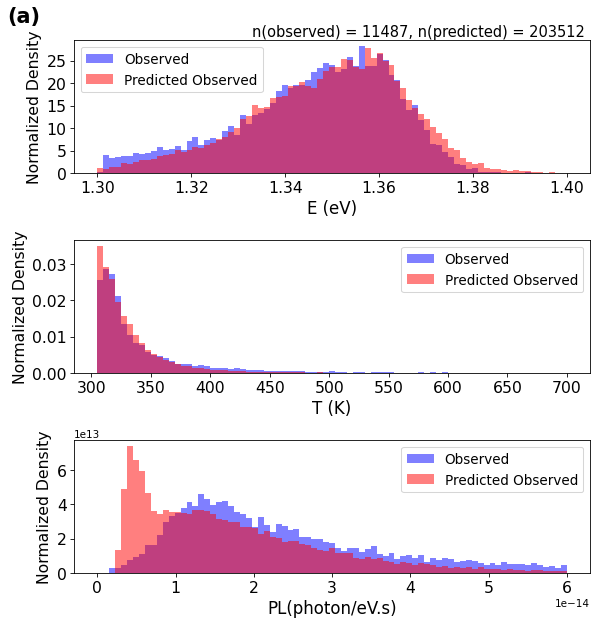}
\includegraphics[width=0.6\textwidth]{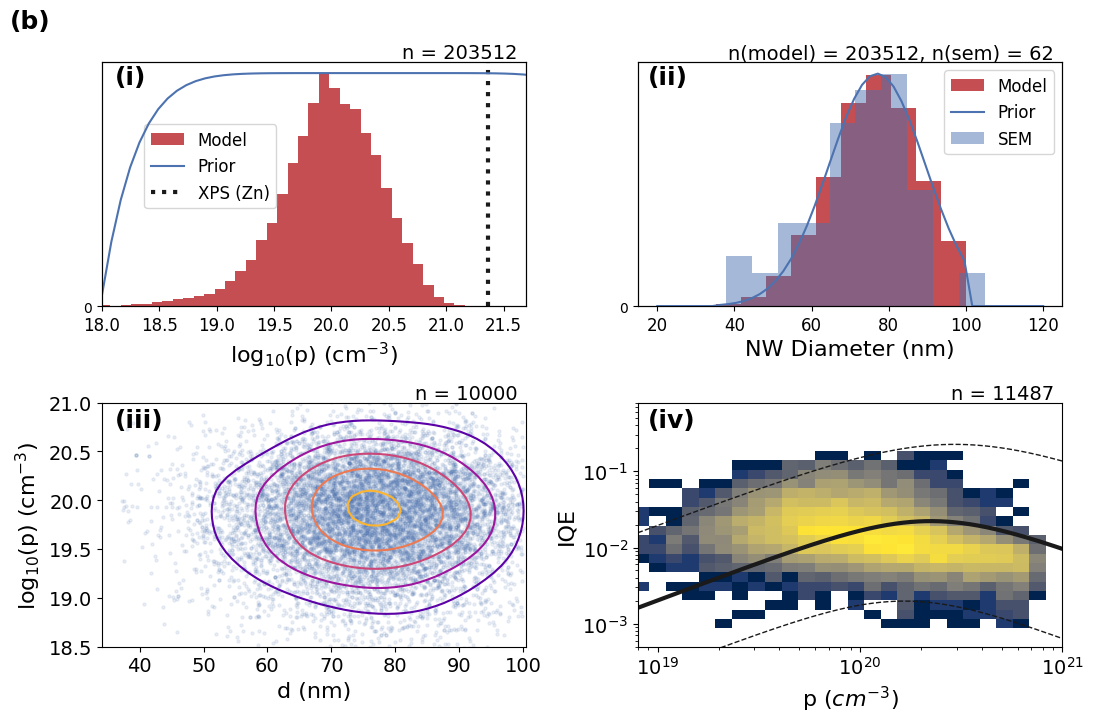}
\caption{(a) Normalized projection of experimentally observed $E_{BE}$, $T_{SO}$ and $PL$ and modified model-predicted $E'$, $T'$, and $PL'$ distributions. \REVISION{The numbers (n) at the top of the figure indicate the number of data points that contribute to each distribution.} (b) Prior distribution (blue lines) and posterior histogram (red) for (i) hole density on log$_{10}$ scale and (ii) NW diameter $d$ with SEM results (blue histogram). The vertical line in (i) shows the Zn density calculated from XPS. (iii) Scatter plot for posterior $p$ and $d$ indicating weak correlation evident from the data. \REVISION{The contours show iso-probability lines calculated using a 2D KDE. Using two-sided Pearson correlation between $\log(p)$ and diameter, the $p$-value$=0.109$ with a linear correlation of $\rho=-0.016$}. (iv) A two-dimensional histogram of model-predicted $IQE$ as a function of $p$, with the black line indicating the modelled $IQE$ with median posterior values, and the dashed lines depict the $IQE$ interquartile range. \REVISION{A correlation calculation results in a $p$-value$\approx0$ and $\rho=-0.304$. The color scale represents increasing density of model outputs, running from blue (low) to yellow (high) density.}}
\label{fig:Distributions}
\end{figure}

As previously noted, by removing data-sets where the signal intensity is too small to be fit, we potentially introduce a bias into our model. While this eliminates known issues associated with clumping and sample contamination, it will also exclude wires with PL below our observable limit, which may be associated with particular regions of doping-diameter space. We can incorporate this into our analysis by assigning a reduced evidential weight ($P_{\rm{evidence}}(\rm{Data}) < 1$) to the regions which are removed during filtering; this approach means that we use the experimental data only to constrain the model within a parameter space where evidence exists. A piecewise evidence function is used, reducing from unity representing unbiased measurement above the 3rd percentile of the experimental distribution, linearly reducing to 0.25 at the 0.5th percentile to capture under-sampling at low emission intensities. While this choice of function is arbitrary, it is found that the modelled results are not highly sensitive to the cutoff values used. Figure \ref{fig:Distributions}a compares one-dimensional projections of ($E',T', PL'$) distributions obtained from our optimized MCMC modelling, showing the experimental results of the full output of our model, and a modified output removing model points below the 1st percentile to more closely match our observed data. The full model distributions tend to reproduce the observed distributions, while there is a slight excess of blueshifted (low-doped) and weak intensity NWs when compared with the experimental results, which is attributed to the bias induced by cutting low-emission NWs and split-off temperature.   

We determine the distribution in doping and NW diameter from the posterior distributions as shown in Figure \ref{fig:Distributions}b(i) and (ii). The posterior distributions determined from sampling versus diameter and doping are illustrated in the SI. From the predicted data, the probability distribution function for hole density is asymmetric with values between \pdoping~\REVISION{which is in line with the previously reported values measured across smaller populations.}\cite{yang2015zn, Johansson2020CalculationNanowires} The value for Zn-doping was obtained from XPS as 1.35(14)$\times10^{21}$\,cm$^{-3}$; while this is significantly higher, it is expected that the Zn would not be fully activated at such high concentrations, and effects such as clustering or interstitial doping may occur. \REVISION{Such shift in doping between XPS and optical method was previously reported for MOCVD-grown Zn-doped GaAs NWs, and calibrated using SIMS and XPS.}\cite{alanis2018optical} \REVISION{We have additionally studied both un-doped and higher-doped GaAs NWs grown using Aerotaxy (data not shown), where we observed that higher-doped NWs contain higher Zn atomic concentrations as measured in XPS, but revealed lower activated hole concentration using optical methods. This supports the existence of segregation and incomplete activation at very high doping levels.} 
The diameter distribution has values of \diameter~which closely follows the prior distribution determined from SEM results.
\REVISION{While optical techniques have been used to measure individual nanowire diameters at this scale, these typically require significant modelling (and therefore prior knowledge), control of polarization, or photo-modulation which is not available in our high-throughput system.}\cite{Bronstrup2010OpticalDevices, Montazeri2011,Chen2020}
The MCMC modelling approach provides likely values for the full vector $\Psi$, which allows us to probe whether the data supports correlations between parameters. Notably, doping and diameter \REVISION{are found to be insignificantly correlated, based on a two-sided Pearson correlation with $p$-value$=0.916$ and $\rho=-0.001$,} as illustrated in Figure \ref{fig:Distributions}b(iii) as a scatter and 2D-Kernel Density Estimate (KDE). This suggests that the dopant incorporation during growth is not primarily dependent on the NW diameter, and other factors are likely to be the dominant source of variation in doping across the NW population. We propose that this is due to the nature of Aerotaxy growth mechanism, where the random growth trajectory of NWs may affect the precursor density or local temperature during growth \REVISION{compared to MBE and MOCVD growth methods.}\cite{kim2021doping,Al-Abri2021_ACS}  

Using our posterior estimates for diameter $d$ and $\alpha$, we are able to produce a model of IQE as a function of doping as shown in Figure \ref{fig:Distributions}b(iv). \REVISION{In particular, we see that the shape of IQE with respect to doping and the variation in IQE at a given doping supports that both doping and other physical properties -- likely diameter -- are important in determining IQE.} It is noted that the model is produced from the median a-posteriori values and Equation~\ref{eqn:pl} and not as a direct fit to the data. While there appears to be an offset between the data and model, we note that the experimental results undersample low-IQE and low-diameter parameter space; indeed there is no requirement for our model to reproduce the data, and the agreement within uncertainty limits is supportive of our approach. We can see two regimes around a switch point $p'=$\pturnp; low doping ($p<p'$): where non-radiative recombination $4S/d$ dominates and IQE increases linearly with doping due to the $Bp$ term in Equation \ref{eqn:pl}, and high doping ($p>p'$): where Auger recombination $Cp^2$ dominates. This is around $4\times$ larger than literature values for this transition determined at 77\,K\cite{Zschauer1969AugerGaAs}; the very large non-radiative rate in narrow GaAs NWs will shift the switching point $p'$ to higher values. The surface recombination velocity $S$ is determined as \SRV, which spans values previously estimated for Zn-doped GaAs NWs with diameter between 50-300\,nm.\cite{darbandi2016measurement, joyce2013electronic, Joyce_2017_ACS} This high $S$ justifies the necessity of control of non-radiative recombination to optimize emission intensity\cite{jiang2012long,jiang2013enhanced}. We highlight that a significant variation in estimated IQE is observed as a function of $p$. This is in-part due to the large spread in NW diameter, and hence relative variation in non-radiative recombination; this can be further explored via the effective emission temperature.

\subsection{Hot Carrier Dynamics}
Our samples demonstrate hot-carrier emission with $T >$ 300\,K, significantly higher than observed in previous studies on thick ($d>$300\,nm) Zn-doped GaAs.\cite{alanis2018optical} Previous time-resolved studies have revealed carrier lifetimes of 1.5\,ps for unpassivated 50\,nm NWs\cite{parkinson2007transient} and 5-7\,ps for 300\,nm Zn-doped NWs\cite{parkinson2007transient, burgess2016doping, Alanis2019}, sufficient time for full thermalization of carriers. However, our present samples have significantly larger radiative and Auger decay channels when compared with undoped 50\,nm NWs and have larger non-radiative surface-related recombination when compared with doped 300\,nm wires; we may naively expect at least a six-fold reduction in lifetime due to the reduction in diameter, to below 1\,ps, giving rise to the hot-carrier emission observed\cite{Wittenbecher2021, Bailey1990NumericalGaAs}. 

An additional complication can arise when interpreting cooling dynamics in degenerately doped semiconductors. Due to the spread in doping, there is a spread in Fermi energy level shift, and with very high doping ($>$\pdeg) GaAs becomes degenerately doped (details in the SI). Photo-excited electrons can then recombine with a population of holes with energy below the valence band-edge, giving rise to a hotter effective emission ($>$1000\,K) than expected from a purely electron-dominated cooling process. Therefore the effect of Fermi temperature spread must be removed from the band-edge temperature to study electron cooling only. By subtracting the non-thermal contribution to emission temperature that arises from degenerate doping the effective electron temperature from the band-edge emission ($T_{BE}$) and split-off emission ($T_{SO}$) become comparable as shown in the SI; a gradient linking these is \slope~indicating that both band-edge and split-off recombination as well as cooling processes are likely to take place at the same timescale. Therefore, the split-off temperature can be exploited as a more accurate ``clock'' for the carrier dynamics. The emission lifetime is linked to the effective electron temperature from the split-off using the median posterior values for $T_0=$\T~and $\tau_0=$\ta~for each of the \usedwires~NW spectra, using
\begin{equation}\label{eqn:tau-T}
    \tau(T) = -\tau_0\ln\left(\frac{T-T_L}{T_0}\right)
\end{equation}

\begin{figure}
\includegraphics[width=0.95\textwidth]{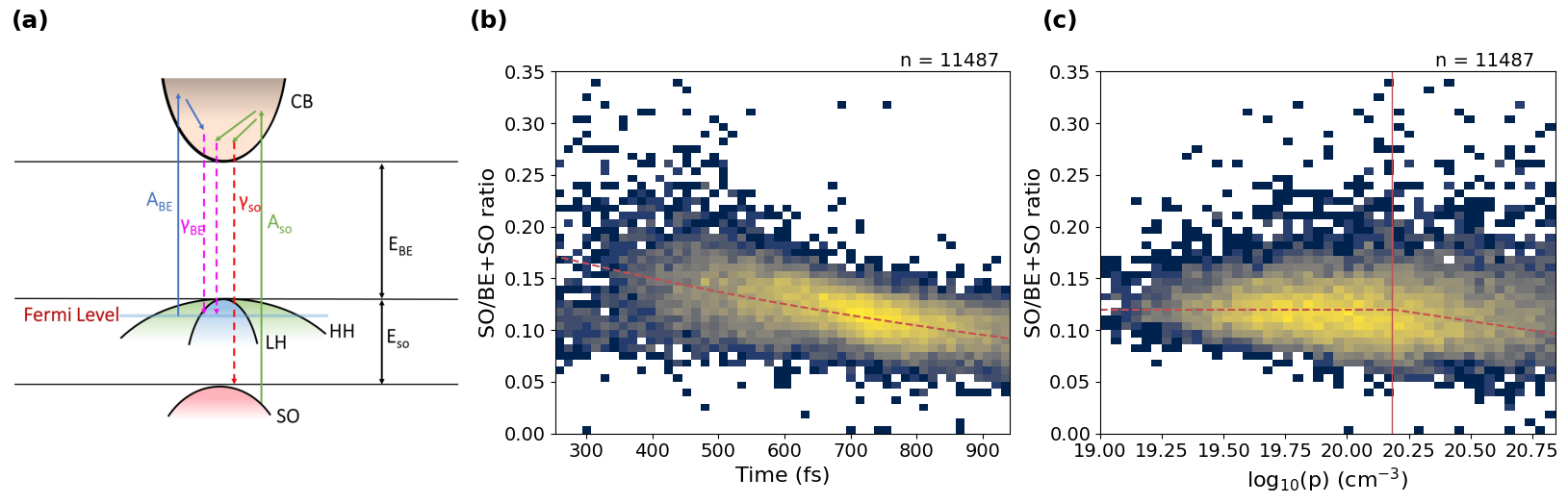}
\caption{(a) A schematic for the carrier dynamics showing pathways of absorption from band-edge $A_{BE}$ and split-off $A_{SO}$ and emission to band-edge $\gamma_{BE}$ and split-off $\gamma_{SO}$. The location of Fermi energy labelled is dependent on doping. (b) Ratio of split-off to the total emission as a function of carrier lifetime shows a short lifetime of \tauFit~from the exponential fit. \REVISION{The number (n) at the top of the figure indicates the number of data points that contribute to the distribution.} The red dotted line is an exponential fit to the data. \REVISION{Two-sided Pearson correlation between split-off to total emission ratio and carrier lifetime shows a $p$-value$\approx$0 and $\rho=-0.515$.} (c) Ratio of split-off to the total emission as a function of logarithmic doping, the vertical line indicates the degenerate doping where $\gamma_{BE}$ dominates and \REVISION{a piece-wise linear fit is provided as a guide to the eye in red. The correlation between emission ratio and doping is $\rho=-0.081$ with $p$-value$\approx$0, based on a two-sided Pearson correlation.}}
\label{fig:TR}
\end{figure}

Figure \ref{fig:TR}a provides a schematic of the proposed dynamic model. Following 1.96\,eV (633nm) photo-excitation, three hole populations will be formed at early times: 40\% in the heavy hole band, 40\% in the light hole band, and 20\% in the split-off band.\cite{Becker1988FemtosecondGaAs} Elsaesser and colleagues showed that at short times following the photo-excitation, split-off emission may be expected to dominate, as excitation from the split-off band can only create carriers in the $\Gamma$ valley, while excitation from the heavy or light hole bands have sufficient excess energy to populate the indirect $L$ or $X$ valleys which reduces their recombination rate.\cite{elsaesser1991initial} The photo-excited electrons in the conduction band have two pathways to recombine; with holes in the band-edge ($\gamma_{BE}$), or with holes in the split-off ($\gamma_{SO}$). To compare these two emission processes, we calculate the ratio of split-off emission to the total emission as a function of recombination lifetime calculated from the temperature of each emission using Equation~\ref{eqn:tau-T} as shown in Figure \ref{fig:TR}b. This ratio is modelled with an exponential decay revealing a short effective carrier lifetime (\tauFit) as expected for thin and heavy-doped GaAs NWs,\cite{parkinson2007transient,burgess2016doping,Alanis2019}, and around 20\% of the total emission from the split-off band which falls with time over a picosecond. 

The rate of reduction in relative split-off emission may be driven and affected by different sub-picosecond processes such as; faster split-off recombination when compared to band-edge recombination owing to stronger coupling as given by Fermi's Golden rule,\cite{ouguzman1995theoretical} hole scattering from the split-off band to the valance band in timescales from 0.35\,ps to 0.5\,ps reducing the split-off band occupation,\cite{Bailey1990NumericalGaAs, Scholz1998HolephononArsenide} or inter-valley scattering from L, X to $\Gamma$ in a timescale between 0.7\,ps to 2\,ps increasing band-edge recombination.\cite{saeta1992intervalley} To give more insight into the full carrier dynamics, additional measurements on thinner nanowires or higher doping density is essential to probe carrier dynamics below 400\,fs. Figure \ref{fig:TR}c shows the ratio of emission processes as a function of calculated doping; here, the average split-off emission is relatively constant at 12\% for non-degenerate doping below \pdeg~(indicated by the red vertical line). However, in the case of degenerate doping the band-edge emission grows due to an increased likelihood of non-geminate recombination to the large light-hole and heavy-hole population, reducing the effective split-off band signal.

\section{Conclusion}
We present a large-scale optoelectronic study of a highly inhomogeneous NW population, performed at a single-wire level for Aerotaxy-grown Zn-doped GaAs NWs. Despite the large spread in doping and diameter obtained from the posterior distribution, our results demonstrate that inter-wire doping inhomogeneity in Aerotaxy-doped GaAs NWs is weakly linked to the NW diameter, suggesting that there are other factors causing doping inhomogeneity. The internal quantum efficiency of the NWs was determined from the surface, radiative, and Auger recombination processes, showing maximum efficiency of \IQEmx~ at doping of \pturnp~when modelled using median a-posteriori parameter values. We show that carrier lifetimes are related to the split-off temperature, where a single process dominates, revealing complex carrier dynamics at a timescales less than \tamx. Our data-driven methodology provides a statistically rigorous evaluation of material properties in the presence of inhomogeneity, as well as a novel approach to studying ultrafast dynamics. High-throughput spectroscopy combined with a Bayesian analytical framework is highly promising for studying bottom-up grown nanomaterials, identifying the origin of inhomogeneity, and providing statistical confidence across a range of material, sample-specific and experimental parameters that are otherwise challenging to access.

\section{\REVISION{Experimental Section}}
\textit{Sample Growth:} 
The NWs investigated in this study were grown using Aerotaxy method following a recipe reported previously\cite{heurlin2012continuous, Sivakumar2020Aerotaxy:Nanostructures}. The grown GaAs NWs are doped with a relative Zn flow rate of 1.5$\%$.

\textit{Micro-Photoluminescence Microscopy:} 
The NWs were characterized using automated micro-PL microscopy. An objective lens was used to focus the laser beam into the sample where the reflected beam from the sample is collected and redirected by non-polarising beamsplitter into a multimode non-polarisation preserving fibre. Hence, the signal received by the spectrometer is effectively polarisation scrambled and no polarisation effects are expected or have been observed. Using a machine vision algorithm,\cite{Church2022HolisticNanowires} the NWs are located and the geometry, orientation, position, and image were acquired for each NW along with a photoluminescence spectra determined. The image processing followed the procedure outlined in Reference \cite{Parkinson2020}, using image-processing tools available in MATLAB (namely $regionprops$).

\textit{Scanning Electron Microscopy:} SEM images were obtained using a Quanta250 FEG microscope. The images were obtained using a secondary electron detector at 2.7\,mm working distance and 3\,keV acceleration voltage. Images were acquired at different magnifications (50$\times$ – 1000$\times$) in which higher magnifications were used to get a good measure for the NW diameter. The NWs geometries from SEM images were obtained using a measurement tool in basic image processing software.

\textit{Statistical Analysis:} 
In our study we have considered 24,819 objects in which the photoluminescence, geometry, orientation, position, and image were obtained. Filtering of the data is based on the removal of weakly-emissive objects (e.g. dust) and very-emissive objects (e.g. clumps) based on the PL emission. Post-filtering was applied after PL fitting based on the physical range of energy, carrier temperature, and emission intensity. This resulted in 11,487 NWs which are used for all the data analysis. Kernel Density Estimation (KDE) is used to convert data points to a probability distribution. A 1D-KDE was used to represent the distribution of length and diameter of NWs, 2D-KDE was used to understand the correlation between doping and diameter and IQE and doping, and 3D-KDE was used to generate 3D space of energy, temperature, and PL distributions. A two-sided Pearson correlation coefficient is used to estimate the strength and direction of the correlation between two variables where $\rho$ gives the correlation coefficient ranging between -1 and 1 (e.g. -1 is direct negative correlation) and $p$-value gives the significance of correlation ranging between 0 and 1 with closer to 0 is more significant correlations.\cite{Akoglu2018UsersCoefficients} For data analysis Jupyter Notebook is used to run Python codes, the code and the data are available online.

\section*{Supporting Information}
Supporting  Information  is  available  from  the  Wiley  Online  Library  or  from the author.

\section*{Acknowledgements}
PP acknowledges funding from UKRI under the Future Leaders Fellowship program (MR/T021519/1). RAA and NAA are in receipt of studentship funding from the Omani Government. Alex Walton acknowledges funding from the EPSRC (UK) (EP/S004335/1). \REVISION{SEM imagery was provided by Dong Wang (University of Manchester)}. MHM acknowledges support from NanoLund, the Swedish Research Council and from the Knut and Alice Wallenberg foundation; the NW growth was done at Lund Nano Lab within the MyFab cleanroom infrastructure.

\section*{Conflict of Interest}
The Authors declare no Competing Financial or Non-Financial Interests.

\section*{Author Contributions}
\textbf{Ruqaiya Al-Abri}: Conceptualization, Methodology, Formal Analysis, Investigation (Optics and SEM), Writing - Original Draft,  
\textbf{Nawal Al Amairi}: Investigation (Absorption),
\textbf{Stephen Church}: Methodology, Formal Analysis, Writing - Review and Editing, 
\textbf{Conor Byrne}: Investigation (XPS), Formal Analysis,
\textbf{Sudhakar Sivakumar}: Resources, 
\textbf{Alex Walton}: Writing - Review and Editing, Supervision, Funding Acquisition,
\textbf{Martin Magnusson}: Writing - Review and Editing, Supervision, Funding Acquisition,
\textbf{Patrick Parkinson}: Conceptualization, Methodology, Formal Analysis, Writing - Review and Editing, Supervision, Project Administration, Funding Acquisition

\section*{Data Availability Statement}
\REVISION{The  data  that  support  the  findings  of  this  study  are  openly  available  in  figshare  at }https://doi.org/10.48420/20025791. The code to perform the analysis is available at \REVISION{TBC}.

\section*{\REVISION{Keywords}}
high-throughput, nanowires, Bayesian, photoluminescence, split-off

\onecolumn
\bibliography{references.bib,ref.bib}

\end{document}


\section{Supporting Information}

\subsection{X-Ray Photoelectron Spectroscopy (XPS) Measurements}
XPS measurements were carried out on the Zn-doped GaAs NWs to determine the atomic concentration of the Zn dopant. The NWs were characterized as-deposited on silicon substrate with a native oxide. X-ray photoelectron spectroscopy (XPS) measurements were performed with SPECS XPS instrument, equipped with a SPECS Focus 500 monochromated Al K$\alpha$ X-ray source with photon energy of 1486.6\,eV and an argon-ion sputtering source. Emitted photoelectrons were collected using a 150\,mm hemispherical energy analyzer  (SPECS Phoibos 150). Detailed scans were recorded for Zn2p and Ga2p core levels at a pass energy of 30\,eV. The areas of the peaks were corrected for the known relative sensitivity factors to calculate Zn:Ga concentration ratios. 

\begin{figure}
\centering
\includegraphics[width=0.7\textwidth]{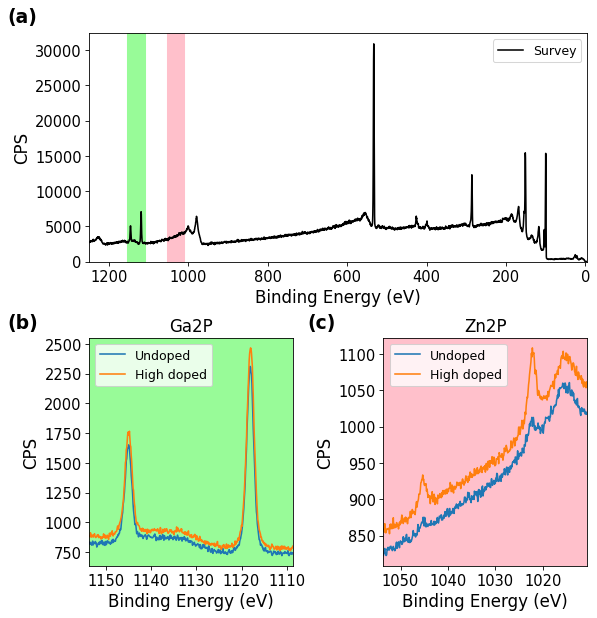}
\caption{(Top) XPS survey spectra for Zn-doped GaAs on silicon oxide substrate. (Bottom) High-resolution XPS scans of the (left) gallium Ga2p and (right) zinc Zn2p regions for both nominally undoped and highly zinc-doped nanowires.}
\label{fig:XPScan}
\end{figure}

The XPS spectrum is shown in Figure \ref{fig:XPScan}a. The sample shows photoemission peaks arising from Ga2p, As3s, and Zn2p which are used to determine the material stoichiometry by dividing the peak areas by their respective relative sensitivity factors for Al K alpha X-Rays. In addition, there are silicon and oxygen peaks associated with silicon oxide substrate and carbon due to the ambient exposure.  Figure \ref{fig:XPScan}b and c show a magnified region at high energy attributed to Ga2P and Zn2p photoemission. Table \ref{tab:sample_table} summarizes the main findings showing nominal Zn flow, the ensemble average of Zn level obtained from XPS, and the effective hole density derived from calculations based on the energy shift reported in the main text. The calculated dopant level using optical methods is noted to be around 20$\times$ lower than that determined from XPS. This may be due to an ensemble weighting effect, or more likely due to incomplete activation of Zn dopants at the high levels present as discussed in the manuscript. 

\begin{table*}[t]
\resizebox{\textwidth}{!}{\begin{tabular}{|l|l|l|l|l|}
\hline
Sample    & Zn Flow    & XPS (Ga:Zn) & XPS Zn density(cm$^{-3}$)    & Optical hole density (cm$^{-3}$) \\ \hline
Undoped-GaAs      & $\sim0$\,\%  & 1:0.012     & 0.26(11)$\times10^{21}$ & -  \\
Zn-GaAs      & 1.5\,$\%$  & 1:0.060     & 1.32(11)$\times10^{21}$ & \pdoping  \\
 \hline
\end{tabular}}
\caption{Summary of reference and Zn-doped GaAs NWs studied showing Zn flow, dopant level $p$ from XPS and hole density from optical measurements from $\mu$-PL spectroscopy.\label{tab:sample_table}}
\end{table*}

\subsection{Post-location Filtering of Nanowires}
There is variation in photoluminescence emission due to inhomogeneity between individual wires, however, SEM imagery (main text) indicates that we also anticipate emission from clumps of multiple NWs and the presence of non-emissive dust or dirt. Following the initial NW location using dark-field optical microscopy, we use a filtering process to remove spectra likely to be contaminants. The photoluminescence emission PL model output is used to create a threshold above which wires are defined as clumps, which are removed before further analyzing. In addition, the emission temperature and peak energy outside reliable range or representing physically unrealistic wires were removed. \REVISION{Examples of the fitting parameters are shown in} Table \ref{tab:fittingParameters} \REVISION{for the four spectra illustrated in the main text.} Table \ref{tab:filtering} shows the benchmarks in which the parameters were filtered and the number of wires removed. This results in removing around 54$\%$ of the wires ending up with \usedwires~wires. 

\begin{table*}[t]
\resizebox{\textwidth}{!}{\begin{tabular}{|l|l|l|l|l|l|l|}
\hline
NW ID & BE energy (eV) & BE temperature (K)  & BE amplitude  & SO energy (eV) & SO temperature (K) & SO amplitude \\ \hline
17469 & 1.331 & 1093 & 1.10$\times10^{-17}$ & 1.76 & 394 & 4.39$\times10^{-18}$ \\
17360 & 1.355 & 1051 & 2.15$\times10^{-17}$ & 1.73 & 385 & 5.75$\times10^{-18}$ \\
13386 & 1.358 & 880  & 6.09$\times10^{-17}$ & 1.73 & 321 & 1.05$\times10^{-17}$ \\
401   & 1.361 & 996  & 9.22$\times10^{-17}$ & 1.72 & 305 & 1.88$\times10^{-17}$ \\

 \hline
\end{tabular}}
\caption{\REVISION{Fitting parameters for the spectra shown in the main text showing peak energy, emission temperature, and emission amplitude for band-edge BE and split-off SO.}\label{tab:fittingParameters}}
\end{table*}

\begin{table*}[t]
\resizebox{\textwidth}{!}{\begin{tabular}{|l|l|l|l|}
\hline
Parameter    & Condition    & Justification & Wires removed \\ \hline
BE emission amplitude & $<6\times10^{-16}$ & Removing clumps & 48 \\
SO emission amplitude & $<1.19\times10^{-16}$ & Removing clumps & 37 \\
BE emission temperature & $<5450$\,K & Removing unphysical high temperatures & 103 \\
BE emission temperature & $>330$\,K & Removing unphysical low temperatures & 962 \\
SO emission temperature & $<700$\,K & Removing unphysical high temperatures & 1928 \\
SO emission temperature & $>305$\,K & Removing unphysical low temperatures & 3216 \\
BE peak energy & $<1.405$\,eV & Removing unphysical high energy & 144 \\
BE peak energy & $>1.301$\,eV & Removing unphysical low energy & 9601 \\
SO peak energy & $<1.779$\,eV & Range of SO emission is known & 1037 \\
SO peak energy & $>1.671$\,eV & Range of SO emission is known & 2893 \\

 \hline
\end{tabular}}
\caption{Summary of filtering conditions of the output PL-model parameters with the number is wires removed. Some wires are removed by multiple filters - the numbers do not sum to the total removed.\label{tab:filtering}}
\end{table*}

\subsection{Scaling of Internal Quantum Efficiency to Photoluminescence Intensity}
The photoluminescence intensity $PL$ is related to the internal quantum efficiency $IQE$ by a scaling factor $\alpha$ and the diameter-dependent NW absorption ($A(d)$) as mentioned in the main text. Quantifying these factors is crucial in setting a prior for Bayesian analysis. 

\subsubsection{Experimental Factor $\alpha$}
The scaling factor $\alpha$ is related to experimental conditions during the acquisition of PL spectra namely laser power, laser spot size, objective lens collection efficiency, microscope throughput efficiency, and spectrometer quantum efficiency. All of these conditions were approximated experimentally by an end-to-end calibration of laser reflection from a mirror in the sample position.

\subsubsection{NW Absorption Modelling}
\begin{figure}
\centering
\includegraphics[width=0.4\textwidth]{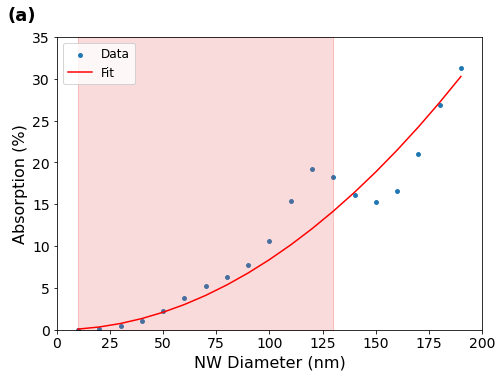}
\includegraphics[width=0.8\textwidth]{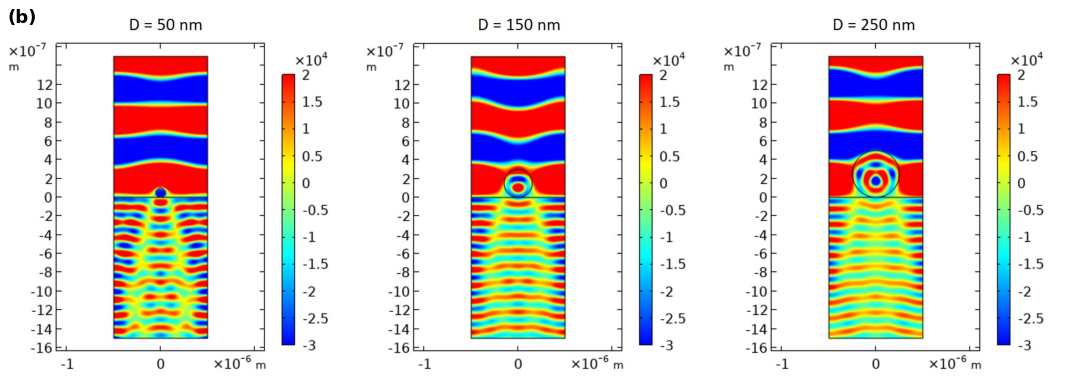}
\caption{(a) Relation between NW absorption of light and diameter, the red line is a second-degree polynomial fit to the data. The shaded area illustrates the range of diameter of interest 10-130\,nm. (b) Electric field distribution displaying light absorption of s-polarized light at three NW diameters 50, 150, and 250\,nm.}
\label{fig:Absorption}
\end{figure}

The NW absorption as a function of diameter was determined using COMSOL simulation. In the simulation, power loss density (PLD) was studied for a range of NW diameters from 10-200\,nm under simulated excitation conditions to obtain the absorption percentage. The PLD was found for p- and s- polarized incident light, with light spot size 1\,$\mu$m. The absorption was obtained by integrating the PLD across the NW length and normalising it to the incident power and spot diameter. Figure \ref{fig:Absorption}a depicts NW absorption of light with second-degree polynomial fit at different NW diameters, the shaded area indicates the range of diameter of interest. While a quadratic fit is naive, reflecting the increasing absorption with increasing geometric area presented to the beam and increasing thickness in the sub-absorption depth regime, the model does not pass through the data points. The NW model used in COMSOL reflects interference related to scattering and substrate interactions, which appear as oscillation. However, the error introduced is relatively small - on the order of $<2\times$ - and we choose to neglect this in our data analysis. An example of NW absorption of light is shown in Figure \ref{fig:Absorption}b for s-polarized light at three different NW diameters.  
  
\subsection{Markov-Chain Monte-Carlo Model}
\subsubsection{Prior Probability Distribution Functions}
\begin{figure}
\centering
\includegraphics[width=0.6\textwidth]{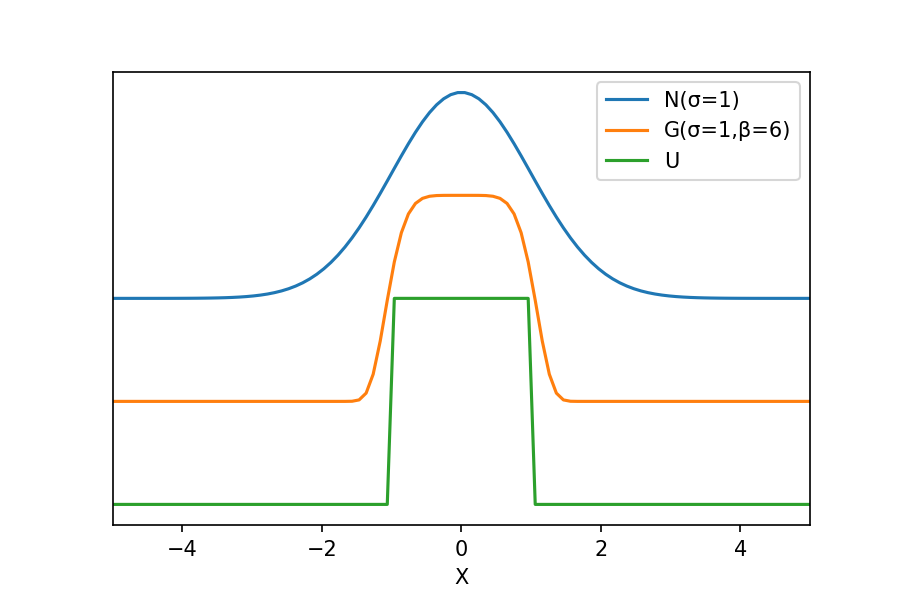}
\caption{Schematic of three probability distribution functions centred at $x=0$ with scale parameter $\sigma=1$.}
\label{fig:PDFs}
\end{figure}

Bayesian inference is a process of updating prior knowledge with new evidence or data. The prior vector $\Psi$ is a list of probability distribution functions (PDFs) that represents our knowledge of each parameter before considering the data. In line with convention, we use analytical PDFs, with our choice of function depending on our confidence in each parameter; these are listed in the main text. The bandgap of GaAs $E_0$ is a well-known material parameter, hence the prior is set as a normal distribution $\mathcal{N}(\mu,\sigma)$ centred at 1.405\,eV with a narrow full width at half maximum (FWHM) of 5\,meV reflecting our small uncertainty. Equation~\ref{equ:normal} describes normal distribution where $\mu$ and $\sigma$ are the mean and standard deviation of the distribution,
\begin{equation} \label{equ:normal}
    \mathcal{N}(\mu,\sigma) = e^{-0.5((x-\mu)/\sigma)^2}.
\end{equation}
In some cases a range of values have been reported in the literature - for instance the radiative recombination coefficient $B$ - or where we base our prior on physical upper and lower limits - for instance the doping level $p$. In this case, a generalized distribution $G(\mu,\sigma,\beta)$ is used, which is a normal distribution with higher uncertainty used 
\begin{equation} \label{equ:generalized}
    \mathcal{G}(\mu,\sigma,\beta) = e^{-0.5((x-\mu)/\sigma)^\beta}.
\end{equation}
The parameter $\beta$ represents the width of the prior distribution. Finally, in some cases hard upper and lower limits are known, such as for carrier initial lifetime. However, we have little further a-priori insight and the prior distribution is therefore uniform 
\begin{equation} \label{equ:uniform}
    \mathcal{U}(x_{min},x_{max}) = (x>=x_{min}) \cap (x<=x_{max}).
\end{equation}
In addition, these distributions may be combined or truncated to be positive to avoid physically unrealistic values. The distributions are schematically shown in Figure \ref{fig:PDFs}.

\subsubsection{Posterior Probability Distributions}
\begin{figure}
\centering
\includegraphics[width=0.45\textwidth]{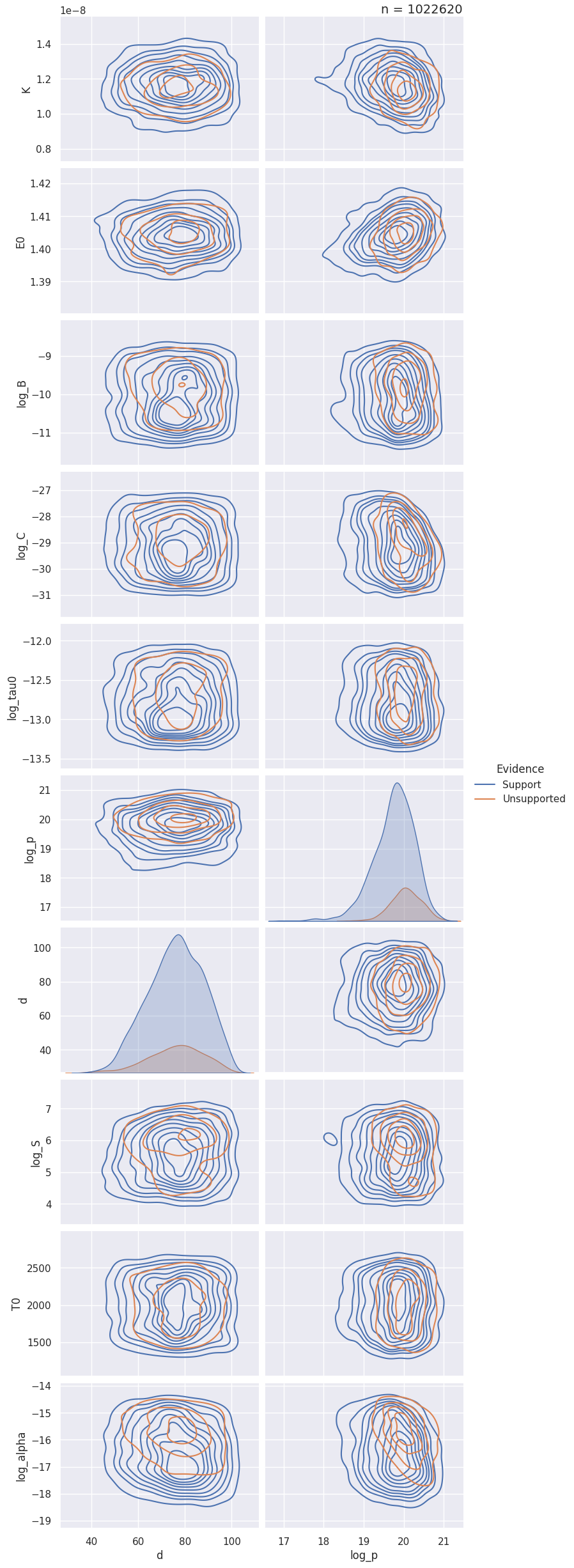}
\caption{Pair posterior correlations for all the parameters determined using MCMC, including $K$, $E_0$, $\log(B)$, $\log(C)$, $\log(\tau_0)$, $\log(p)$, $d$, $\log(S)$, $T_0$, and $\log(\alpha$) versus $d$ and $\log(p)$. \REVISION{The number (n) at the top of the figure indicates the number of samples under investigation.}}
\label{fig:Posteriors}
\end{figure}

Figure \ref{fig:Posteriors} shows the distributions and correlation plots between all model posterior distributions as a function of doping $\log(p)$ and diameter $d$. As described in the main text, posterior samples are separated into regions with experimental support (evidenced regions) and the full range returned by the model. Additionally, samples with very high values of $\alpha$ are removed to improve the visualization of the correlations ($>$\alphaSampling) resulting in removing $7\%$ of the sampling data-set. The figure illustrates no strong correlations between the parameters as a function of $d$ and $\log(p)$ before and after masking; the correlation between $\log(p)$ and $d$ is investigated in detail in the main text. 

\subsection{Fermi Energy Shift}
\begin{figure}
\centering
\includegraphics[width=0.95\textwidth]{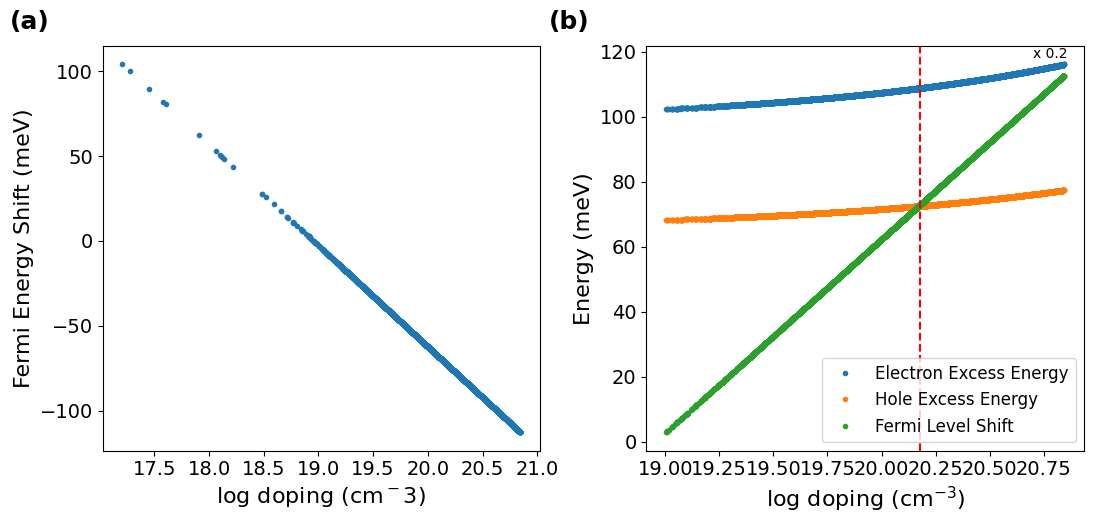}
\caption{(a) Fermi energy shift as a function of logarithmic doping. (b) Hole excess energy and scaled electron excess energy versus doping compared to Fermi energy. The vertical line indicates the doping level beyond which GaAs NWs absorption into the valance band is affected.}
\label{fig:Eexcc_Ef}
\end{figure}

The location of the Fermi level in a p-type doping semiconductor is proportional to the logarithmic hole density, as given in the following relation
\begin{equation} \label{equ:Fermi}
    \Delta E = K_{B}T \ln\left(\frac{N_{V}}{p}\right),
\end{equation}
where $\Delta E$ is the location of the Fermi level with respect to the valence band-edge, $K_{B}T$ is thermal energy given as 25\,meV, $N_{V}$ is the density of state in the valence band given as $9\times10^{18}$\,cm$^{-3}$, and $p$ is the hole density. Given the spread in hole density, the shift of Fermi level is calculated for each doping measured as illustrated in Figure \ref{fig:Eexcc_Ef}a. The negative values indicate that Fermi level is below the valence band. For a Fermi-level within the gap, no correction is needed. However, the doping regime above $1\times10^{19}$\,cm$^{-3}$ leads to degenerate doping, and a correction to the effective emission temperature is required. The electron excess energy $E_{exc}^{e}$ and hole excess energy $E_{exc}^{h}$ due to the incident photo-excitation energy $E_{inc}$ = 1.96\,eV are calculated using the hole-electron effective mass ratio $r_{m^{*}}$ = \meff~ and the spread in the band-edge energy $E_{g}$, that is
\begin{align} \label{equ:Excess}
    E_{exc} = E_{inc}-E_{g} \\
    E_{exc}^{h} = E_{exc}\times r_{m*} \\
    E_{exc}^{e} = E_{exc}-E_{exc}^{h}
\end{align}
Figure \ref{fig:Eexcc_Ef}b depicts the Fermi energy, electron, and hole excess energies with respect to doping. Fermi energy intercepts the hole excess energy at doping \pdeg~revealing that above this doping level GaAs NWs absorption to the valance band will drop.      

\subsubsection{Effective Carrier Temperature}
\begin{figure}
\centering
\includegraphics[width=0.65\textwidth]{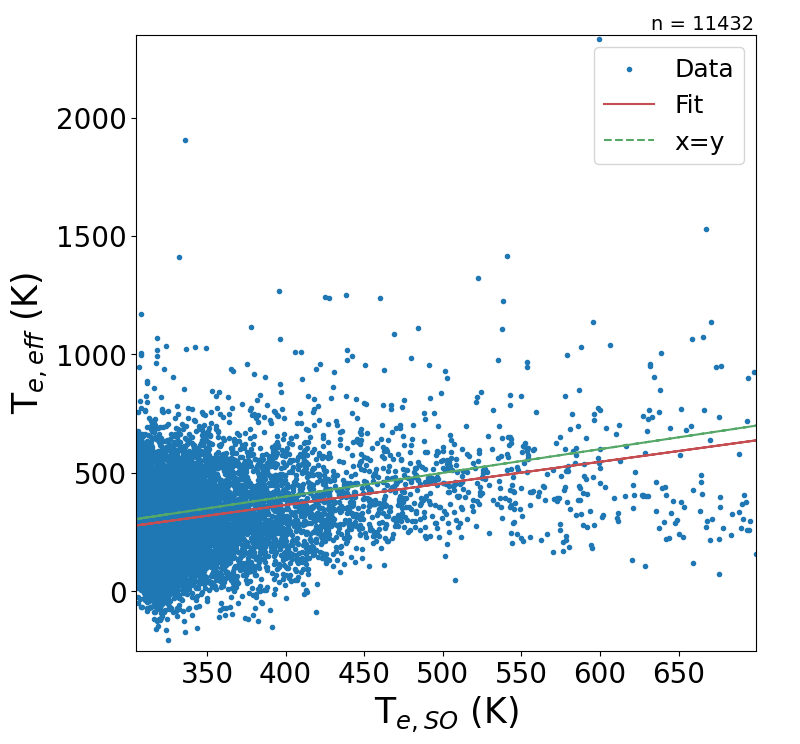}
\caption{Comparison of effective electron temperature from the band-edge and the split-off temperature\REVISION{, at positive Fermi energies,} showing a gradient of \slope K/K~. \REVISION{The number (n) at the top of the figure indicates the number of samples under investigation. Using two-sided Pearson correlation between effective electron temperature from the band-edge and the split-off temperature with $p-value=0$ and $\rho=0.268$}.}
\label{fig:Effective_Temp}
\end{figure}

Due to the spread in Fermi energy level shift, the photo-excited electrons can recombine with holes below the valence band edge, giving rise to hot emission. By eliminating the effect of Fermi temperature, the effective electron temperature from the band-edge and split-off are comparable as shown in Figure \ref{fig:Effective_Temp}; with a gradient of \slope\,K/K. This indicates that both band-edge and split-off recombination processes are likely to take place at the same timescale. \REVISION{Here we show only a high doping regime i.e., positive Fermi energy.}

